\documentclass[conference]{IEEEtran}
\IEEEoverridecommandlockouts
\usepackage{amsmath,amssymb,amsfonts}
\usepackage{algorithmic}
\usepackage{graphicx}
\usepackage{booktabs}
\usepackage{multirow}
\usepackage{subfig}
\usepackage{float}
\usepackage[section]{placeins}
\usepackage{cite}
\usepackage{makecell}
\usepackage{textcomp}
\usepackage{xcolor}
\usepackage[ruled,linesnumbered]{algorithm2e}
\DeclareRobustCommand*{\IEEEauthorrefmark}[1]{%
    \raisebox{0pt}[0pt][0pt]{\textsuperscript{\footnotesize\ensuremath{#1}}}}
\def\BibTeX{{\rm B\kern-.05em{\sc i\kern-.025em b}\kern-.08em
    T\kern-.1667em\lower.7ex\hbox{E}\kern-.125emX}}
\begin{document}

\title{RIS-aided Real-time Beam Tracking for a Mobile User via Bayesian Optimization}

\author{
\IEEEauthorblockN{
Junshuo Liu\IEEEauthorrefmark{1}, Rujing Xiong\IEEEauthorrefmark{1}, Jialong Lu\IEEEauthorrefmark{1}, Tiebin Mi\IEEEauthorrefmark{1}, 
Robert Caiming Qiu\IEEEauthorrefmark{1}}
\IEEEauthorblockA{\IEEEauthorrefmark{1}School of Electronic Information and Communications, Huazhong University of Science and Technology, Wuhan 430074, China}
Emails: \{junshuo\_liu, rujing, M202272434, mitiebin, caiming\}@hust.edu.cn
}

\maketitle

\begin{abstract}
The conventional beam management procedure mandates that the user equipment (UE) periodically measure the received signal reference power (RSRP) and transmit these measurements to the base station (BS). The challenge lies in balancing the number of beams used: it should be large enough to identify high-RSRP beams but small enough to minimize reporting overhead. This paper investigates this essential performance-versus-overhead trade-off using Bayesian optimization. The proposed approach represents the first application of real-time beam tracking via Bayesian optimization in RIS-assisted communication systems. Simulation results validate the effectiveness of this scheme.
\end{abstract}


\section{Introduction}
Recent studies have emphasized the potential of reconfigurable intelligent surfaces (RIS) in reconfiguring wireless propagation environments by adjusting the phase shifts of passive reflecting elements, offering cost-effectiveness and low power consumption\cite{wu2019towards,shojaeifard2022mimo,liu2022path}. RIS arrays can modulate amplitude and phase to create specific reflected beams, enhancing signal power and suppressing interference in desired directions. Dynamic beam tracking, a key challenge, involves designing suitable reflection coefficients since the user's position is non-deterministic. Three primary approaches for beam tracking are channel state information (CSI)-based, side-data-assisted, and codebook-based schemes.

The CSI-based approach has received considerable attention in recent research. For instance, the user mobility is assumed to follow the first-order Gaussian-Markov model, and an Extended Kalman filter (EKF) method is used to track the optimal beam\cite{zegrar2020general}. It is the first work that addresses channel estimation, beamforming, and user tracking under millimeter wave (mmWave) RIS-aided multiple-input-multiple-output (MIMO) systems. Similarly, the works \cite{liu2023channel,zhang2022ris} rely on the assumption that path coefficients evolve over time and a Kalman filter is used to track the primary channel direction. Additionally, Reference \cite{yu2023kalman} applies the Kalman filter to exploit temporal correlation for tracking cascaded channels.

In side-data-assisted schemes, reference \cite{ouyang2023computer} introduced a computer vision-based approach to aid RIS for dynamic beam tracking, along with the implementation of a prototype verification system. The experimental results demonstrated that RIS could rapidly adapt reflection coefficients for dynamic beam tracking with the assistance of visual information. The incorporation of out-of-band information into communication systems served to reduce beam training overhead and eliminate feedback links.

Codebook-based schemes, in contrast to CSI-based approaches, offer simplicity as they don't require intricate channel estimation, making them advantageous for RIS systems with numerous array elements. One such codebook-based passive RIS configuration method was proposed to avoid complex channel estimation\cite{wang2021codebook}. In the work\cite{kaya2021deep}, they assume UE mobility follows predictable patterns and predict future optimal beams based on prior received signal reference power (RSRP) measurements using a long short-term memory (LSTM) deep learning algorithm. The primary challenge here is the training phase, where the BS accumulates an extensive set of UE reports and trains the LSTM model.

Bayesian optimization (BO) is a valuable method for optimizing complex functions when facing costly evaluations, lack of derivatives, or non-convex problems\cite{shahriari2015taking}. BO employs a Gaussian process (GP) model to predict function values and uncertainties. Based on these predictions, BO selects input points likely to yield optimal results before performing physical measurements. Motivated by the Bayesian optimization algorithm, this paper introduces a novel approach using Bayesian optimization for real-time beam tracking of mobile users with RIS. The main contributions of this paper can be summarized as follows:
\begin{itemize}
\item[$\bullet$] For the first time, an effective Bayesian optimization-based approach is proposed to assist RIS in realizing real-time beam tracking. The RIS plays a crucial role in enhancing the signal coverage and determining the codebook based on the BS and UE positions.
\item[$\bullet$]The proposed approach is evaluated under two UE mobility speeds, demonstrating the BS's capability to rapidly adjust reflection coefficients for dynamic beam tracking without relying on historical data, enabling real-time beam tracking for a mobile user.
\end{itemize}

\textbf{Notations.} The imaginary unit is denoted as $j$, and the magnitude of a complex number is represented by $\left | \, \cdot \, \right |$. Bold lowercase and uppercase letters represent vectors and matrices, respectively. The conjugate transpose of $\boldsymbol{A}$ is denoted as $\boldsymbol{A}^H$.

\section{Bayesian Optimization: Preliminaries}
Consider the following problem of finding a global maximizer (or minimizer) of an unknown objective function
\begin{equation}
\boldsymbol{x}^{*}=\arg \max_{\boldsymbol{x}\in \mathcal{X}} f(\boldsymbol{x})\label{bo}
\end{equation}
where $\mathcal{X}$ is the compact set of the domain, i.e., a hyper-rectangle $\mathcal{X} = { \{\boldsymbol{x} \in \mathbb{R}^{D}: a_{i} \leq x_{i} \leq b_{i} \}}$, where $D$ is the dimension of $\boldsymbol{x}$. In Bayesian optimization, a probabilistic model for the objective function $f(\boldsymbol{x})$ is used to inform the selection of evaluation points within the domain $\mathcal{X}$. This approach maintains a historical record $\mathcal{H}_t=\{ (\boldsymbol{x}_i, f(\boldsymbol{x}_i)), i = 1, 2, ..., T \}$ of previously observed objective values and estimates a surrogate model $\mathcal{M}_t$ based on this history. The next parameters for evaluation are determined by optimizing an acquisition function $\mathcal{A}(\boldsymbol{x}_t, \mathcal{M}_{t-1})$ using information from the surrogate distribution over the parameter space $D$.

Bayesian optimization employs surrogate models like Gaussian Processes and Tree-structured Parzen Estimator (TPE)\cite{bergstra2011algorithms} for estimation. It selects new sampling points by minimizing an acquisition function, such as Probability of Improvement (PI), Expected Improvement (EI), or Upper Confidence Bounds (UCB).

\begin{algorithm}
\caption{The pseudo-code for Bayesian optimization}\label{algorithm1}
$\mathcal{H}_0 \leftarrow \emptyset$\;
  \For{$i \leftarrow 1$ \KwTo $T$}{
    $x_t \leftarrow \arg \max_{x\in D} \mathcal{A}(x, \mathcal{M}_{t-1})$\;
    $\mathrm{Evaluate}$ $f(x_t)$\;
    $\mathcal{H}_t \leftarrow \mathcal{H}_{t-1}$ $\cup$ $(x_t, f(x_t))$\;
    $\mathcal{M}_t \leftarrow $ $\mathrm{update}$ $\mathrm{Surrogate}$ $(\mathcal{H}_t, \mathcal{M}_{t-1})$\;
    }
${x}^{*}, f(x^{*}) \leftarrow $ $\mathrm{find}$ $\mathrm{Maximum} (\mathcal{H}_{T})$\;
\Return ${x}^{*}, f(x^{*})$
\end{algorithm}

\textbf{Surrogate model.} The GP provides a powerful prior distribution for functions, denoted as $f: \mathcal{X} \rightarrow \mathbb{R}$. It is defined by the property that a finite set of $S$ points $\{\boldsymbol{x} \in \mathcal{X} \}_{n=1}^S$ induces a multivariate Gaussian distribution in $\mathbb{R}^S$. The characteristics of this distribution for functions are determined by a mean function $m: \mathcal{X} \rightarrow \mathbb{R}$ and a positive definite covariance function $K: \mathcal{X} \times \mathcal{X} \rightarrow \mathbb{R}$. For instance, the radial basis function (RBF) kernel is:
\begin{equation}
k_{\theta}^{\mathrm{RBF}}(x,x')=\theta_1 \exp{ \left( \frac{\delta(x,x')}{\theta_2^2} \right)} \label{rbf}
\end{equation}
where $\delta$ is a distance metric and $\theta = [\theta_1, \theta_2]$ is the vector of hyper-parameters. The kernel determines the smoothness of function $f$ with respect to the metric $\delta$.

In contrast to Gaussian process-based modeling of $p(y|x)$, the TPE approach models $p(x|y)$ and $p(y)$. To achieve this, TPE transforms the generative process by replacing configuration prior distributions with non-parametric densities. By incorporating different observations $\{ x^{(1)}, ..., x^{(k)}\}$ in these densities, it creates a versatile learning algorithm that can represent various densities across the configuration space $\mathcal{X}$. TPE defines $p(x|y)$ using two such densities:
\begin{equation}
p(x|y) = 
\begin{cases}
l(x), & \mbox{if } y < y^* \\
g(x), & \mbox{if } y \ge y^*
\end{cases}\label{tpe}
\end{equation}
where $y^*$ represents the best value found after observing $\mathcal{H}: y^* = \max \{ f(x_i),1\leq i \leq T\}$, $l(x)$ is the density using observations $\{ x^{(i)}\}$ that corresponding loss $f(x^{(i)})$ was less than $y^*$, and $g(x)$ is the density using the remaining observations. The approach further sets $y^*$ to a quantile $\gamma$ of the observed $y$ values, ensuring $p(y<y^*)=\gamma$.

\textbf{Acquisition function.} The acquisition function balances exploration and exploitation within the objective space $\mathcal{X}$ for optimal $x$. We choose Expected Improvement due to its excellent performance. EI is defined as:
\begin{equation}
EI_{y^*}(x) = \int_{-\infty}^{+\infty} \max (y^*-y, 0) \, p_{\mathcal{M}}(y|x)\, dy \label{ei}
\end{equation}
where $y^*$ serves as a threshold. Expected Improvement measures the likelihood of $f(x)$ exceeding (negatively) $y^*$, given $x$. The hyper-parameter $x_t$ with the highest EI is identified as the local optimal hyper-parameter. After evaluating $f(x_t)$, BO stores both $x_t$ and $f(x_t)$ in the search history, updates the model $\mathcal{M}_t$, and, upon completion of the iterative process, provides the global optimal hyper-parameter.

Combining Eqs. (\ref{tpe}), (\ref{ei}), optimization of EI in the TPE approach is concluded to be:
\begin{equation}
    \begin{aligned}
EI_{y^*}(x) &= \int_{-\infty}^{y^*} (y^*-y) \, p(y|x)\, dy\\  
            &= \int_{-\infty}^{y^*} (y^*-y) \frac{p(x|y)p(y)}{p(x)} \, dy \\
            &= \int_{-\infty}^{y^*} (y^*-y) \frac{l(x)p(y)}{\gamma l(x) + (1 - \gamma)g(x)} \, dy \\
            &= \frac{\int_{-\infty}^{y^*} (y^*-y)p(y) \, dy}{\gamma + (1-\gamma)\frac{g(x)}{l(x)}}
            \label{5}
    \end{aligned}
\end{equation}

Therefore, $EI_{y^*}(x)=\frac{\gamma y^* l(x)-l(x)\int_{-\infty}^{y^*} p(y)dy}{\gamma l(x)+(1-\gamma) g(x)} \propto (\gamma + \frac{g(x)}{l(x)}(1-\gamma))^{-1}$. This expression emphasizes the preference for points $x$ with high probability under $l(x)$ and low probability under $g(x)$ to maximize improvement. The tree-structured format of $l$ and $g$ simplifies candidate generation based on $l$ and their evaluation via the $g(x) / l(x)$ ratio. The algorithm selects the candidate $x^*$ with the highest EI in each iteration.

By combining both the surrogate model and the acquisition function, we can
now perform Bayesian optimization in Algorithm \ref{algorithm1}. BO is an iterative process with three main components. At each iteration, we first infer the reward at unmeasured points via the GP model. Then, we pick a new point to measure. Finally, we tune the GP model.

\textbf{Inference.} Up to iteration $i - 1$, we have selected points $\mathbf{x}_{i-1} = \{ x_1,...,x_{i-1} \}$ and observed their rewards $\widetilde{\mathbf{f}}_{i-1} = \{ \widetilde{f}(x_1),...,\widetilde{f}(x_{i-1}) \}$. At iteration $i$, we aim to infer the reward $f(x)$ for any point $x$. The GP framework assures that the random variables $\widetilde{\mathbf{f}}_{i-1}$ and $f(x)$ are jointly Gaussian. This allows us to infer $f(x)$ from prior measurements using the Gaussian posterior probability.

\textbf{Choice of next point.} Selecting the next point $x_i$ typically involves maximizing an acquisition function $\mathcal{A}$ that addresses the exploration-exploitation dilemma. We seek to exploit past observations by choosing $x_i$ where the GP posterior mean is high, while also exploring uncharted areas of $\mathcal{X}$ where the GP standard deviation is high. A well-known acquisition function is expected improvement, which measures the expected reward enhancement when selecting $x$ over the highest expected reward.

The difference between the TPE model and the GP model lies in the fact that TPE provides more detailed modeling of historical observed results. It segments and models the observed results in sections, allowing for a segmented probability distribution modeling.

\section{System Description}
In this section, we present the system model, outline the optimization problem for RIS-assisted beam tracking, and introduce the design principle for the phase configuration matrix (codebook). For the sake of clarity, we'll use the terms "phase configuration matrix" and "codebook" interchangeably in the subsequent discussion.

\begin{figure}[tbp]
  \centering
  \includegraphics[width=0.95\columnwidth]{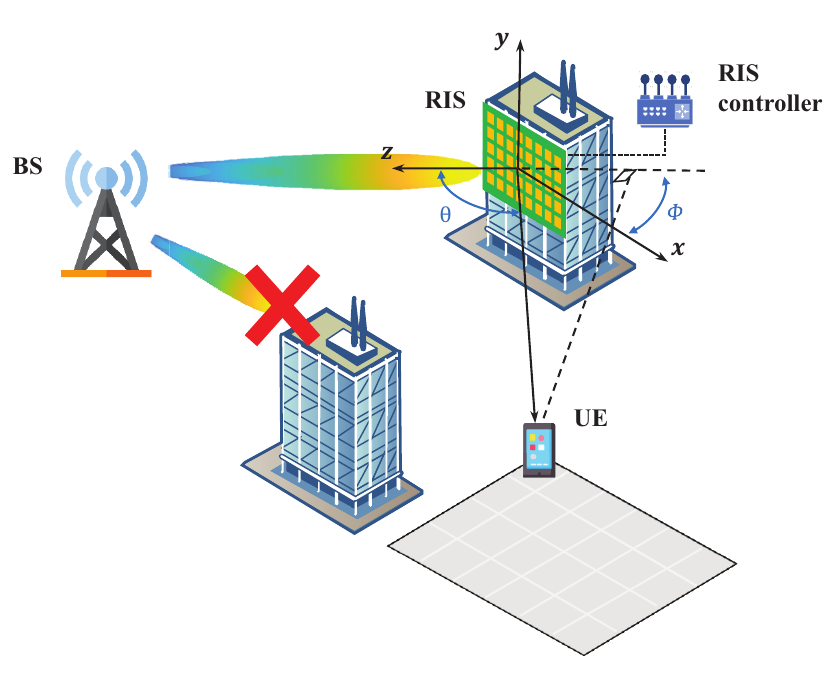}
  \caption{The RIS-assisted beam tracking system.}\label{F1}
\end{figure}
\subsection{System model}
We consider a base station with $M$ antennas serving a single-antenna user with the assistance of a RIS comprising $N$ passive reflecting elements, as shown in Fig. \ref{F1}. The RIS is connected to a smart controller that dynamically configures the phase shift of RIS by the BS. Assumptions include blocked direct links between the BS and UE due to obstacles, with the UE moving at low to medium speeds within the cell. Time is divided into slots, and during each slot, the transmit signal at the BS is expressed as $ \boldsymbol{z}_t \in \mathbb{C}^{M}$. The RIS receives the signal as
\begin{equation}
\boldsymbol{r}_t = \boldsymbol{H}_t \boldsymbol{z}_t
\end{equation}
where $\mathbf{H}_t \in \mathbb{C}^{N \times M}$ represents the BS-to-RIS channel. Upon receiving $\boldsymbol{r}_t$, the RIS applies precoding using $\boldsymbol{W}$, where $\boldsymbol{W} = \mathrm{diag}(e^{j\beta_1}, e^{j\beta_2},...,e^{j\beta_N}) \in \mathbb{C}^{N\times N}$. Here, $e^{j\beta_n}$ signifies the reflection coefficient and $\beta_n \in [-\pi, \pi]$ is the phase shift at the $n$-th RIS antenna element ($n=1, 2, ..., N$). The signal received at the UE is expressed as
\begin{equation}
y_t = \boldsymbol{h}^{H}_t \boldsymbol{W} \boldsymbol{H}_t \boldsymbol{z}_t + n_t
\end{equation}
where $\boldsymbol{h}^{H}_t \in \mathbb{C}^{N}$ denotes the RIS-to-UE channel, and $n_t$ is the additive Gaussian white noise.

Considering the path loss for the BS-to-RIS and RIS-to-UE channels, these channels can be modeled as
\begin{equation}
\boldsymbol{H}_t [i, k] = \frac{\lambda}{4 \pi d_{1,ik}} e^{-j \frac{2\pi d_{1,ik}}{\lambda}}
\end{equation}
\begin{equation}
\boldsymbol{h}^{H}_t [i] = \frac{\lambda}{4 \pi d_{2,i}} e^{-j \frac{2\pi d_{2,i}}{\lambda}}
\end{equation}
where $d_{1,ik}$ denotes the distance between $i$-th element of RIS and $k$-th antennae of BS, $d_{2,i}$ represents the distance between $i$-th element of RIS and the UE, and $\lambda$ is the wavelength.

\subsection{Problem formulations}
We aim to find a phase configuration matrix to maximize the UE's RSRP, which can be formulated as:
\begin{equation}
\max_{\boldsymbol{W}} f(\boldsymbol{W}) \label{eq10}
\end{equation}
Since the channel $\boldsymbol{h}^{H}_t$ and $\boldsymbol{H}_t$ are unknown, the problem in (\ref{eq10}) cannot be solved by the conventional optimization method.

One simple approach to mitigate this issue is for the BS to select any phase configuration matrix $\boldsymbol{W}$ and request the UE to measure and report $| y_t |^2$ values. The UE then employs the phase configuration matrix $\boldsymbol{W}$ that yields the highest $| y_t |^2$ for data transmission. However, this process incurs significant feedback overhead when the UE must measure numerous phase configurations. Thus, the tracking problem aims to select an appropriate phase configuration matrix $\boldsymbol{W}_{\mathrm{opt}}$ that balances the achieved $| y_t |^2$ performance with the feedback overhead.

\subsection{RIS and codebook design} \label{codebooks}
The RIS consists of 100 units arranged in a 10$\times$10 grid, each spaced at half-wavelength intervals. Each unit has a metal patch, a bias line, two PIN diodes, and ground. By adjusting the bias voltage across the PIN diode, two states are achieved: "1" for the forward bias state and "0" for the reverse bias state. The digital control module manipulates bias voltages to create four different discrete phase coding states: "00" (0{$^\circ$}), "01" (90{$^\circ$}), "10" (180{$^\circ$}), and "11" (270{$^\circ$}). 

For coordinates, we use $(x_{rx}, y_{rx}, z_{rx})$ for the UE and $(x_{tx}, y_{tx}, z_{tx})$ for the BS. Pitch angle $\theta_{rx}$ and azimuth angle $\phi_{rx}$ describe the UE's orientation relative to the RIS. Direction $(\theta_{rx}, \phi_{rx})$ of the UE, given $\theta_{rx} \in [0, \frac{\pi}{2}]$ and $\phi_{rx} \in [0, 2\pi]$, can be calculated as follows:
\begin{equation}
\theta_{rx} = \arctan\frac{\sqrt{x_{rx}^2 + y_{rx}^2}}{z_{rx}}
\end{equation}
\begin{equation}
\phi_{rx} = \arctan\frac{y_{rx}}{x_{rx}}
\end{equation}
Based on the 3D coordinates of the BS and the direction of $(\theta_{rx}, \phi_{rx})$ of the UE, we can obtain the phase $\beta_{mn}$ of the incident beam on each RIS unit, allowing us to establish the corresponding codebook. We utilize the divide-and-sort (DaS) algorithm to calculate the optimal reflection coefficients for specific directions\cite{xiong2022optimal}. Hence, the codebook datasets can be pre-computed and solidified.

\section{Bayesian optimization for beam tracking}
Bayesian optimization, owing to its predictive capabilities and improved estimation accuracy with fewer measurements, is a well-suited choice for real-time beam tracking. In this section, we present the design principles, evaluation metrics, and the proposed algorithm for beam tracking.

\subsection{Design principles}
1) Upon a UE's entry into the cell, the BS's goal is to generate a set of codebooks for efficient tracking of high RSRP beams while minimizing tracking cycles. This strategy is driven by the aim to optimize data rate transmission for the UE.

\begin{figure}
  \centering
  \includegraphics[width=0.95\columnwidth]{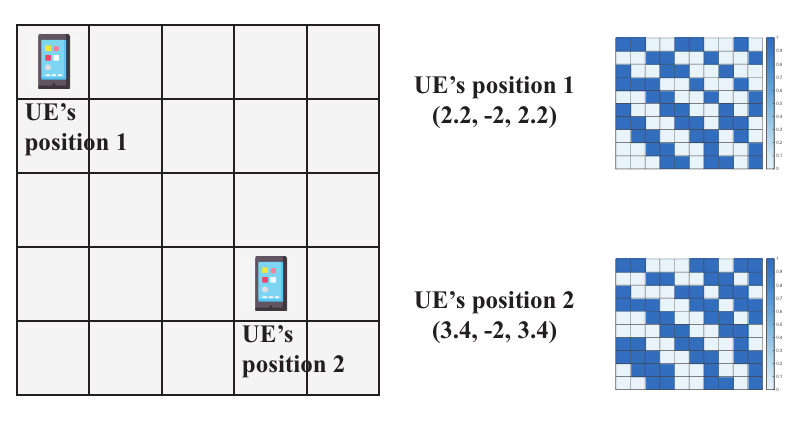}
  \caption{UE positions correspond to phase configuration matrices.}\label{F2}
\end{figure}

2) Effective tracking without measuring all phase configurations is possible due to the correlations in RSRP across the different phase configuration matrices. Fig. \ref{F1} shows the cell divided into $100$ areas, each a square of $0.4\times0.4\, \mathrm{m}^2$. These areas correspond to unique phase configuration matrices $\boldsymbol{W}_{\mathrm{all}}$, benefiting from the good spatial resolution provided by the RIS. Thus, as depicted in Fig. \ref{F2}, user positions on the left grid and the corresponding RIS codebooks on the right.

3) The BS must adapt the quantity of proposed phase configuration matrices to the UE at each iteration, reducing them as RSRP uncertainty decreases.

\subsection{Evaluation metrics}
In this study, we employ BO for phase configuration matrix selection and evaluate performance based on: i) Accuracy, representing the probability of $| \boldsymbol{h}^{H}_t \boldsymbol{W}_t \boldsymbol{H}_t \boldsymbol{z}_t |^2 = \max_{\boldsymbol{W} \in \boldsymbol{W}_{\mathrm{all}}} | \boldsymbol{h}^{H}_t \boldsymbol{W} \boldsymbol{H}_t \boldsymbol{z}_t |^2$; ii) Overhead, indicating the proportion of phase configuration matrices used in BO compared to the total; iii) Average RSRP Error, quantifying the mean absolute error (MAE) between $| \boldsymbol{h}^{H}_t \boldsymbol{W}_t \boldsymbol{H}_t \boldsymbol{z}_t |^2$ and $\max_{\boldsymbol{W} \in \boldsymbol{W}_{\mathrm{all}}} | \boldsymbol{h}^{H}_t \boldsymbol{W} \boldsymbol{H}_t \boldsymbol{z}_t |^2$; iv) Execution Time, denoting the total algorithm execution time.

\subsection{Beam tracking via Bayesian optimization}

The total training procedure consists of two stages: the initial stage and the updating stage. First, in the initialization process, the BS, the RIS, and the UE randomly initialize the phase configuration matrix $\boldsymbol{W}_t$. Then, the UE estimates the RSRP $y_t$ and feeds it back to the BS. The BS fits a GP based on the selected phase configuration matrix $\boldsymbol{W}_t$ and its corresponding RSRP $y_t$. Subsequently, it selects the next phase configuration matrix $\boldsymbol{W}_{t+1}$ for updating the GP based on EI. Finally, by repeating this updating process, the RSRP is maximized. The proposed Bayesian optimization-based training scheme is shown in Algorithm \ref{algorithm2}.

\begin{algorithm}
\caption{Bayesian optimization for user tracking}\label{algorithm2}
\textbf{Initialization.} Choose randomly a codebook from the pre-computed codebooks in Sect. \ref{codebooks}\;
UE computes the RSRP according to the selected codebook\;
UE connects to BS at iteration $i=0$\;
  \For{$i \leftarrow 1$ \KwTo $n$}{
    BS fits a GP model based on the selected codebooks and corresponding RSRP values\;
    BS uses the EI to select the next codebook\;
    }
\end{algorithm}

\section{Simulation results}

In this section, we present numerical results to verify the performance of these approaches. Numerical results are obtained using MATLAB, and the results are averaged over $100$ independent optimization epochs. The total number of time slots is $T = 12$ $\mathrm{s}$. We consider a three-dimensional (3D) scene where the BS and RIS are located at $(0, 0, 0.5)$ and $(0, 0, 0)$, respectively. In addition, the UE is located at a $4\, \mathrm{m} \times 4\, \mathrm{m}$ rectangle area and moves in each $0.4 \times 0.4 \, \mathrm{m}^2$ grid at a selected speed. The path loss models from the BS to RIS and from RIS to UE are $11 + 2 \log_{10}{(d (\mathrm{m}))}$, where $d$ denotes the distance. The noise power $\sigma^2$ is set as $-120$ dBm. The simulation specifications are found in Table \ref{table1}.
\begin{table}[htbp]
\centering
\caption{Simulation configuration parameters.}
\label{table1}
\begin{tabular}{lll}
\toprule
Parameters            & Symbol        & Value                         \\ \midrule
Operating frequency   & $f_c$         & $5.8$ $\mathrm{GHz}$          \\
RIS dimensions        & $N$           & $10 \times 10$                \\
RIS element distance  & $d_{\lambda}$ & $0.0259$ $\mathrm{m}$         \\
Wavelength            & $\lambda$     & $0.0517$ $\mathrm{m}$         \\
Light speed           & $c$           & $3\times 10^8$ $\mathrm{m/s}$ \\
Number of BS antennas & $M$           & $2$                           \\
Noise power           & $\sigma^2$    & $-120$ $\mathrm{dBm}$         \\
UE speed              & $s$           & $1, 2$ $\mathrm{grid/s}$   \\
Time slot             & $T$           & $12$ $\mathrm{s}$             \\
\bottomrule
\end{tabular}
\end{table}

We compare the ergodic algorithm, a GPy implementation of Gaussian process regression (GPR) with an RBF kernel, and a HyperOpt implementation of TPE-based BO\cite{bergstra2013making}. For these methods, we first sample some phase configuration matrices $\boldsymbol{W}_{\eta} \in \boldsymbol{W}_{\mathrm{all}}$, where the subscript $\eta$ denotes the proportion of $\boldsymbol{W}_{\mathrm{all}}$, and choose the best phase configuration matrix $\boldsymbol{W}_{\mathrm{opt}}$ according to $\arg \max_{\boldsymbol{W} \in \boldsymbol{W}_{\mathrm{all}}} f (\boldsymbol{W})$. We consider three sampling fractions: $\eta \in \{0.2, 0.4, 0.6\}$.

\begin{table*}[tbp]
\centering
\caption{Performance of three methods at varying UE speeds.}
\label{table2}
\begin{tabular}{cccccccc}
\toprule
\multirow{2}{*}{Model}  & \multirow{2}{*}{Overhead} & \multicolumn{3}{c}{$s = 1 \, \mathrm{grid/s}$} & \multicolumn{3}{c}{$s = 2 \, \mathrm{grid/s}$} \\ \cline{3-8} 
                        & & Accuracy & RSRP error & Execution time (s) & Accuracy & RSRP error & Execution time (s) \\ 
                        \hline
             Ergodic    & 1.0 & 1.000 & 0.000 & 0.335 & 1.000 & 0.000 & 0.339   \\ 
                        \hline
\multirow{3}{*}{GPR}    & 0.2 & 0.200 & 11.715 & 0.168 & 0.196 & 6.877 & 0.142  \\ 
                        & 0.4 & 0.400 & 5.974  & 0.259 & 0.408 & 2.911 & 0.324  \\
                        & 0.6 & 0.583 & 3.171  & 0.343 & 0.597 & 1.754 & 0.346  \\ \hline
\multirow{3}{*}{TPE-BO} & 0.2 & 0.556 & 3.090 & 0.0914 & 0.558 & 1.533 & 0.0845 \\ 
                        & 0.4 & 0.748 & 1.650 & 0.173  & 0.752 & 1.209 & 0.172  \\
                        & 0.6 & 0.916 & 0.517 & 0.261  & 0.925 & 0.209 & 0.251   \\
\bottomrule
\end{tabular}
\end{table*}

The performance of all schemes for two different UE speeds is presented in Table \ref{table2}. The ergodic algorithm achieves 100\% accuracy with 0 RSRP error, but it comes at the cost of lengthy sweep time and additional feedback links, leading to increased communication delay. Particularly when dealing with a large codebook dataset, the ergodic algorithm's execution time becomes impractical for real-time tracking. In contrast, the TPE-BO algorithm excels in real-time beam tracking, requiring only 0.0914 s with the same overhead, making it a superior choice compared to the GPR and ergodic algorithms.

\begin{figure}[htbp]
  \centering
  \includegraphics[width=0.95\columnwidth]{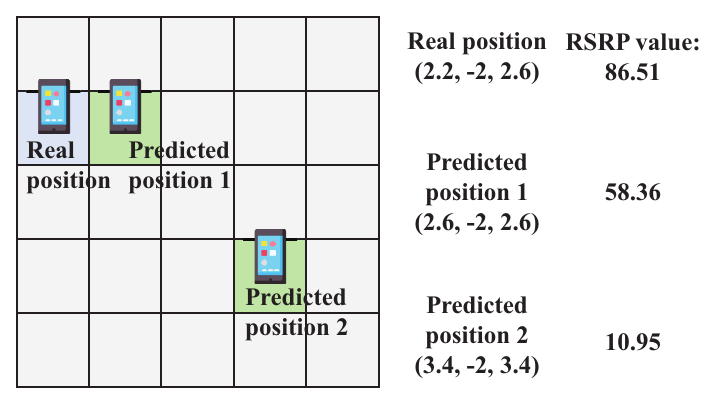}
  \caption{The relationship between the positions and RSRP values}\label{F3}
\end{figure}

The GPR falls short of achieving acceptable accuracy and RSRP error, even with an overhead of up to 0.6. In contrast, the TPE-BO method excels when the overhead reaches 0.6, demonstrating an accuracy of 91.6\% and a low RSRP error of 0.517. Even with an overhead of 0.4, the TPE-BO method attains a 74.8\% accuracy and a 1.650 RSRP error at a UE speed of 1 grid/s. As illustrated in Fig. \ref{F3}, substantial prediction deviations correspond to significant RSRP errors, while smaller prediction errors correlate with more accurate UE positions. For instance, the predicted position \#2 is far from the real position, resulting in a large difference between the two RSRP values.

\begin{figure}[tbp]
    \centering
    \subfloat[Overhead = 0.4]{
		\includegraphics[width=.97\columnwidth]{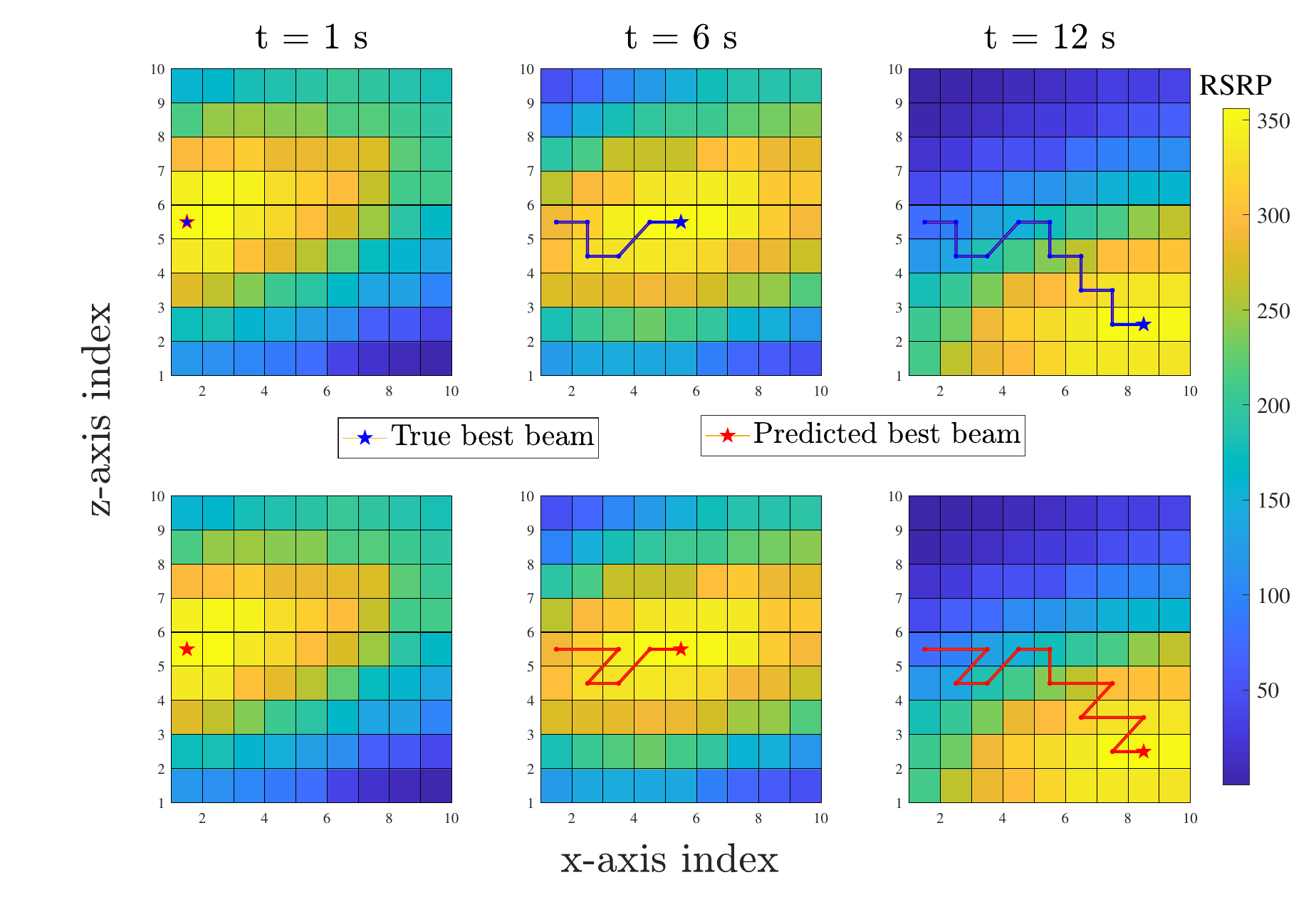}}
    \\
    \subfloat[Overhead = 0.6]{
		\includegraphics[width=.97\columnwidth]{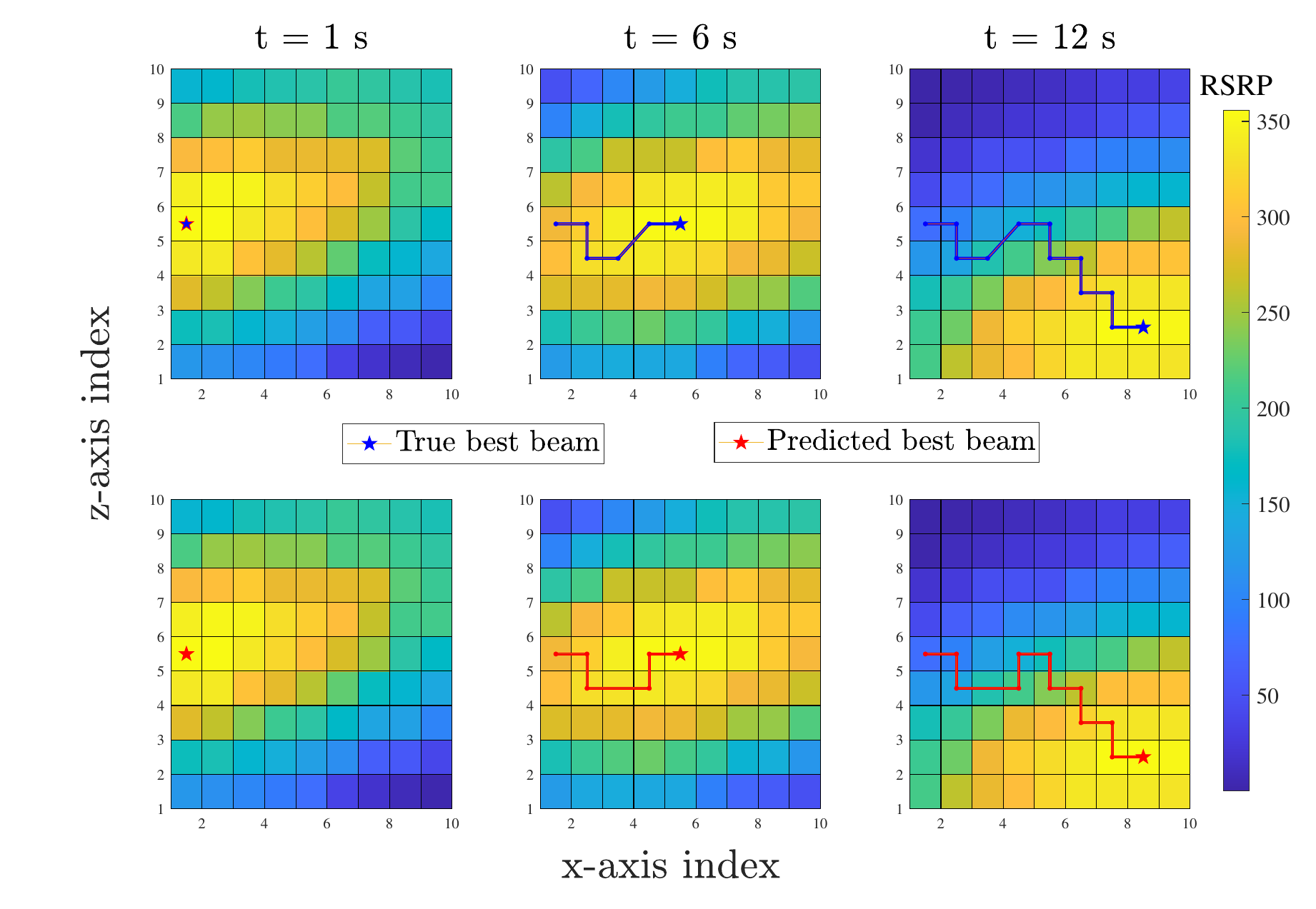}}
    \caption{Time history of a typical UE at 0 s, 6 s, and 12 s. Top row: True RSRP landscape (blue star) and path trace (blue line). Bottom row: Predicted best beam index (red star) and path trace (red line).}
    \label{F4}
\end{figure}

In Fig. \ref{F4}, we present a typical UE time history, visualizing the predicted UE path trace alongside the ground truth. The TPE-BO algorithm, with its small RSRP error, leads to only minor deviations in predicted user positions from the true positions at time instants t = 2 s, 8 s, and 10 s.

Table \ref{table2} presents the performance of these methods under various user speeds. It's evident that they excel in the RSRP error when the user speed is 2 grid/s. This is mainly due to the spatial resolution limitations of the RIS, which, in our simulations, had a 10$\times$10 size. Due to the relatively small size of the RIS panel, the beams formed by the RIS are not narrow enough, resulting in less concentrated energy. Consequently, RSRP differences between closely spaced points are less distinct, challenging the BO algorithm's performance. However, with higher UE speeds, the increased spacing between sampling points results in more pronounced RSRP variations, enabling better fitting and improved performance.

\section{Conclusion}
In this paper, we have demonstrated how Bayesian optimization offers an effective approach for beam tracking, allowing the UE to maintain a connection to a high-RSRP beam by measuring a limited number of codebooks per time slot. Simulation results confirm the algorithm's capability to achieve real-time beam tracking using pre-computed codebooks and stabilize the RSRP of the UE. There are several potential avenues for future work. First, implementing a prototype system would further validate the Bayesian optimization algorithm's effectiveness. Second, exploring the impact of the spatial resolution provided by RIS on beam tracking is also a worthwhile research direction.

\section*{Acknowledgment}
This work was supported by the Nation Natural Science Foundation of China under Grant No.12141107.

\end{document}